\documentclass{aa}
\usepackage{graphicx}
\usepackage{color}
\usepackage{txfonts}
\usepackage{lscape}
\begin{document}

   \authorrunning{Li et al.}

   \title{Conversion from mutual helicity to self-helicity observed with IRIS}

   \author{L. P. Li \inst{1,2}, H. Peter\inst{2}, F. Chen\inst{2},
   and J. Zhang \inst{1}}

   \institute{Key Laboratory of Solar Activity, National Astronomical
   Observatories, Chinese Academy of Sciences, 100012 Beijing, China\\
              \email{lepingli@nao.cas.cn}
         \and
             Max Planck Institute for Solar System Research (MPS), 37077
             G\"ottingen, Germany\\}
   \date{Received ????; accepted ????}

% \abstract{}{}{}{}{}
% 5 {} token are mandatory

  \abstract
   {In the upper atmosphere of the Sun observations show convincing evidence for
   crossing and twisted structures, which are interpreted as mutual helicity and self-helicity. }
   {We use observations with the new Interface Region Imaging Spectrograph (IRIS)
   to show the conversion of mutual helicity into self-helicity in coronal structures
   on the Sun.}
   {Using far UV spectra and slit-jaw images from IRIS and  coronal images and
   magnetograms from SDO, we investigated the evolution of two crossing loops in
   an active region, in particular, the properties of the Si IV line profile in cool loops.}
   {In the early stage two cool loops cross each other and accordingly have mutual
   helicity. The Doppler shifts in the loops indicate that they wind around
   each other. As a consequence, near the crossing point of the loops (interchange)
   reconnection sets in, which heats the plasma. This is consistent with the
   observed increase of the line width and of the appearance of the loops at higher
   temperatures. After this interaction, the two new loops run in parallel, and in one
   of them shows a clear spectral tilt of the Si IV line profile. This is indicative
   of a helical (twisting) motion, which is the same as to say that the loop has self-helicity.}
   {The high spatial and spectral resolution of IRIS allowed us to see the conversion
   of mutual helicity to self-helicity in the (interchange) reconnection of
   two loops. This is observational evidence for earlier
   theoretical speculations. }


   \maketitle
%
%________________________________________________________________

\section{Introduction}

Magnetic helicity is a key quantity that characterizes the complexity of
a magnetic configuration in terms of topology and of the linkage of
magnetic field lines. When a magnetic structure is
built up by two (or more) substructures, one can introduce the
concept of self-helicity and mutual helicity (Berger
\cite{berger99}). Self-helicity corresponds to the twist and
writhe of confined bundles of magnetic flux, while mutual
helicity characterizes the crossing of field lines in the magnetic
configuration (R\'egnier et al. \cite{regnier05}). Using TRACE
images, Chae (\cite{chae00}) determined the mutual helicity of
filaments based on two crossing threads. For the active region (AR)
NOAA 8210, R\'egnier et al. (\cite{regnier05}) reported that the
magnetic configuration was dominated by mutual helicity and
not by self-helicity. With theoretical arguments, Berger
(\cite{berger99}) speculated that mutual helicity can be converted
into self-helicity if two magnetic loops reconnect. Georgoulis
(\cite{georgoulis11}) proposed that numerous small-scale magnetic
reconnection events can lead to an effective transformation of mutual
into self-magnetic helicity. Tziotziou et al. (\cite{tziotziou13})
found a hysteresis in the buildup of self-helicity with respect to
mutual helicity, and considered this as a possible conversion of
mutual- into self-helicity. They also suggested that magnetic
reconnection is the way to achieve this. However, as of today, there
is no observational confirmation for the conversion from mutual into
self-helicity or vice versa.

One way to detect self-helicity is to investigate the spectral tilt
of an emission line. In the image of a spectral line in a slit
spectrometer (i.e., space vs. wavelength), a static bright feature
appears as a line along the wavelength direction that is
perpendicular to the slit direction. However, if the structure shows
a spinning motion, one side of the feature will show a redshift, the
other a blueshift. This leads to a tilt of the line, so that the
bright feature in the spectral image is no longer perpendicular to
the slit direction. In spectral images this is easy to see and is
often called spectral tilt. The detection of this spinning motion is
similar to the detection of a spectroscopic binary.

The spinning nature of spicular features was found by investigating
the spectral tilt in H$\alpha$ spectra (Pasachoff et al.
\cite{pasachoff68}; Rompolt \cite{rompolt75}). Cook et al.
(\cite{cook84}) observed the spectral tilt in transition region
emission features in C IV spectra. Pike \& Mason (\cite{pike98})
reported blueshift and redshift on opposite sides of macrospicules
and interpreted them as a rotating motion. Tian et al.
(\cite{tian08}) reported both red and blue Doppler shifts in bright
points and proposed that this might result from a twist of the
associated magnetic loop system. Kamio et al. (\cite{kamio10})
detected redshift and blueshift on two sides of a macrospicule and
explained this with the unfolding motion of a twisted magnetic flux
rope. Curdt \& Tian (\cite{curdt11}) reported simultaneous Doppler
flows of a red and a blue component without apparent motion and
considered this as spectroscopic evidence for helicity in explosive
events. Based on observations of bifurcated structures and spectral
tilt, Curdt et al. (\cite{curdt12}) proposed that there are spinning
motions in transition region jets. De Pontieu et al.
(\cite{depontieu12}) detected a spectral line tilt in off-limb
spicules, which they considered to be the signature of torsional
motion. Orozco Su\'arez et al. (\cite{orozco12}) found opposite
Doppler shifts at the edge of a prominence foot and interpreted
these shifts as prominence plasma that rotates around the axis.
Recently, Wedemeyer et al. (\cite{wedemeyer13}) and Su et al.
(\cite{su14}) separately found redshifted and blueshifted regions on
the two sides of the prominence leg and proposed that there is
rotational motion in a tornado-like prominence.

\begin{figure*}
   \centering
   \sidecaption
   \includegraphics[width=12cm]{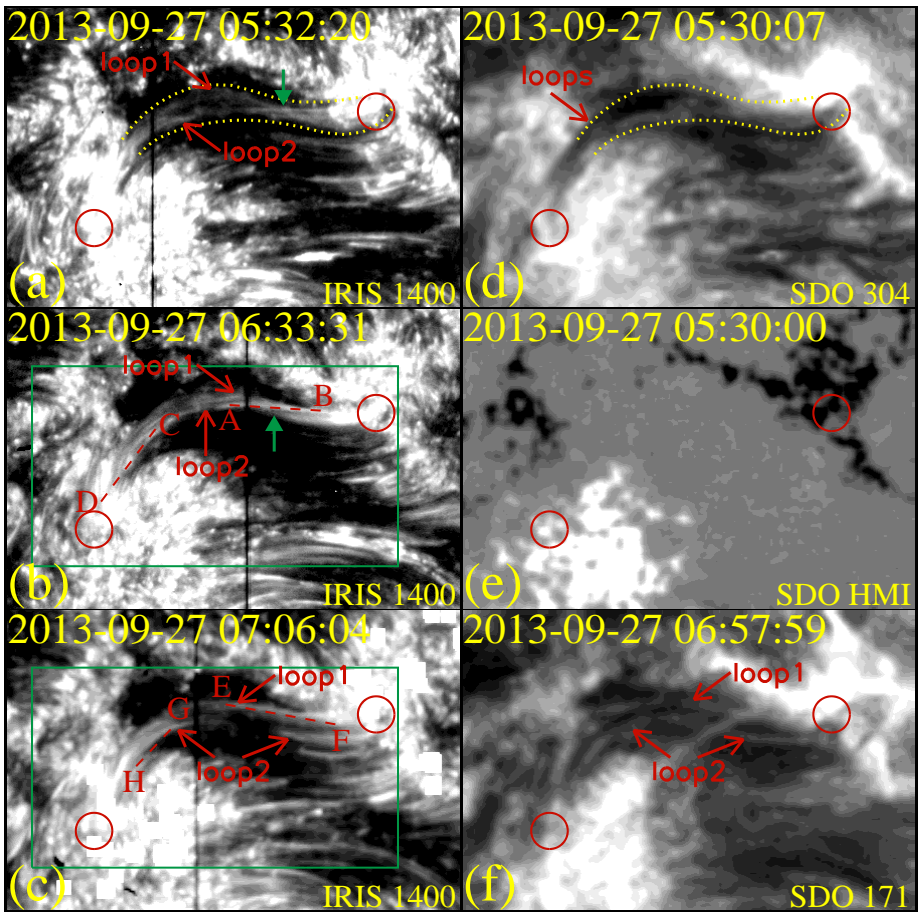}
   \caption{IRIS slit-jaw images and context from AIA and HMI.
    The left panels (a)-(c) show snapshots of the IRIS slit-jaw images
    during the three raster scans. The right panels show the SDO
    context: the AIA 304\,\AA\ image (d) and
    the HMI LOS magnetogram (e) during the first scan, and
    the AIA 171\,\AA\ image (f) during the third scan. The yellow dotted  lines
    in panels (a) and (d) outline the envelope containing the two cool loops seen
    in the IRIS slit-jaw images.
    The red circles mark the endpoints of these loops. The green
    arrows in (a) and (b) indicate the crossing point.
    The red dashed lines
    in (b) and (c) show the positions for time-space
    diagrams displayed in Fig. \ref{F:timeslices}.
    The green rectangles in (b) and
    (c) are the FOV of Figs. \ref{F:spectrographs2} and
    \ref{F:spectrographs3}.
    The center of this image is at solar $(x,y)=(319\arcsec,102\arcsec)$,
    and the FOV is
    75$\arcsec$$\times$50$\arcsec$. The temporal evolution of the IRIS slit-jaw
    images at 1400\,\AA\  of the AIA 304\,\AA\ and 171\,\AA\ channels the HMI
    magnetograms is available in the online
    edition.}
    \label{F:images}%
\end{figure*}

All these observations together show that spinning and twisting
motions are common on the Sun. New observations with the Interface
Region Imaging Spectrograph (IRIS; De Pontieu et al.
\cite{depontieu14a}) now clearly show that these torsional motions
related to self-helicity are a ubiquitous feature of the solar
atmosphere (De Pontieu et al. \cite{depontieu14b}). With the common
observations of mutual helicity, that is, crossing of loops,
threads, and so on, the question arises whether these two forms of
helicity are converted into each other, as speculated by Berger
(\cite{berger99}).

To determine whether there is a helicity conversion, we mainly
employed spectra and images of the transition region acquired by
IRIS. Adding context data of the corona and the photospheric
magnetic field, we investigated the change of the mutual and
self-helicity of two cool loops in an active region. Our results
clearly indicate that mutual helicity is converted into
self-helicity.

\section{Observations and data processing}\label{S:obs}

The IRIS observatory provides simultaneous images and spectra of the
photosphere, chromosphere, transition region, and corona (De Pontieu
et al. \cite{depontieu14a}). From 05:00 UT to 08:00 UT on September
27, 2013, IRIS observed the active region AR 11850 and acquired three
raster scans with the slit spectrograph. The two loops we
concentrate on here are located to the north of the AR. As
we show below, the first and second observing sequences were
made during a phase of mutual helicity, the third raster during
a self-helicity phase. We used IRIS level 2 data, which are already
corrected for flatfield, geometric distortions, and dark
current\footnote{The IRIS data are available at
http://iris.lmsal.com/data.html.}. The general information on the
observations is listed in Table \ref{table:1}. For the first and
second scans, large dense raster modes are employed with steps
of 0.35\arcsec\ . For the third  scan, a coarse raster mode with
steps of 2\arcsec\ is used. During the first two scans, all available
wavelength bands are included in the slit-jaw images (SJI), that
is, 1330\,\AA, 1400\,\AA, 2796\,\AA,\ and 2832\,\AA. In the third scan,
only  images in the 1400\,\AA\ band are recorded. We scaled the maps
and images to the same pixel scale for all scans and co-aligned all
IRIS data according to the information in Table \ref{table:1}.
Figures \ref{F:images}a-\ref{F:images}c separately show the IRIS
1400\,\AA~band\,SJI images during the three observing sequences.

\begin{table*}
\caption{General information of IRIS raster scans and slit-jaw images.} %
\label{table:1}      % is used to refer this table in the text
\centering                          % used for centering table
\begin{tabular}{c|c c c c c c }        % centered columns (4 columns)
\hline\hline                 % inserts double horizontal lines
 & & & Spectrograph raster scans & & & \\
\hline
  & Center & Field & Spatial sampling & Step & Exposure & Steps \\

[UT] & [\arcsec] & of view [\arcsec] & along slit [\arcsec/pixel] & cadence
[s] & time [s] & \\% table heading
\hline                        % inserts single horizontal line
05:24:32-05:48:21 & N322.4,W71.1 & 140.5$\times$180.5 & 0.166 & 3.6
& 2 & 400$\times$0.35\arcsec \\ % inserting body of the table

06:24:43-06:44:10 & N329.8,W63.9 & 140.5$\times$182.3 & 0.166 & 2.9
& 2 & 400$\times$0.35\arcsec \\ % inserting body of the table

06:59:49-07:33:08 & N357.9,W76.2 & 126.8$\times$123.4 & 0.333 & 31.2
& 30 & 64$\times$2.01\arcsec \\
\hline
 & & & Slit-jaw images (1400\,\AA) & & & \\
\hline
  & Center & Field & Spatial sampling & Time & Exposure & \\

 [UT] & [\arcsec] & of view [\arcsec] & [\arcsec/pixel] & cadence [s] & time
[s] & \\
\hline
05:24:32-05:48:21 & N329.5,W70.7 & 309.7$\times$180.5 & 0.166 & 12 & 2 \\
% inserting body of the table

06:24:43-06:44:10 & N330.7,W63.6 & 312.7$\times$182.3 & 0.166 & 12 & 2 \\
% inserting body of the table

06:59:49-07:33:08 & N360.3,W75.9 & 244.5$\times$123.4 & 0.333 & 31 & 30 \\
\hline                                  %inserts single line
\end{tabular}
\end{table*}

To place the IRIS data in the context of the structure and evolution
from the photosphere to the corona we used data from the Solar
Dynamics Observatory (SDO; Pesnell et al. \cite{pesnell12}), in
particular, of the Atmospheric Imaging Assembly (AIA; Lemen et al.
\cite{lemen12}) and the Helioseismic and Magnetic Imager (HMI; Schou
et al. \cite{schou12}). The time cadence and spatial sampling of the
AIA images are 12 s and 0.6\arcsec/pixel. For the HMI line-of-sight
(LOS) magnetograms the respective numbers are 45 s and
0.5\arcsec/pixel. We scaled the SDO observations to match the IRIS
SJI and co-aligned them using several characteristic features, such
as network and plage patterns. Figures
\ref{F:images}d-\ref{F:images}f show an AIA 304\,\AA\ image, an HMI
magnetogram, and an AIA 171\,\AA\ image taken during the IRIS
observations. A movie showing the temporal evolution of the Sun as
seen in the IRIS and SDO data is available online.

IRIS scans the whole structure for the second and third raster
scans. During the first raster scan the IRIS spectral data only
cover the left half of the field of view (FOV) shown in
Fig.\,\ref{F:images}, however. Therefore, we mainly used the spectra
of the second and third raster scans to investigate the properties
of these two loops. For the spectroscopic analysis we used the
transition region lines of Si\,IV at 1394\,\AA\ and 1403\,\AA. For
the first two scans we used the 1394\,\AA\ line because it is
stronger. During the third raster scan only the  1403\,\AA\ line has
been recorded. The Si\,IV profiles were approximated by
single-Gaussian fits with a continuum to build maps of intensity,
Doppler shift, and line width.

\begin{figure}
   \centering
%   \sidecaption
   \includegraphics[width=8.8cm]{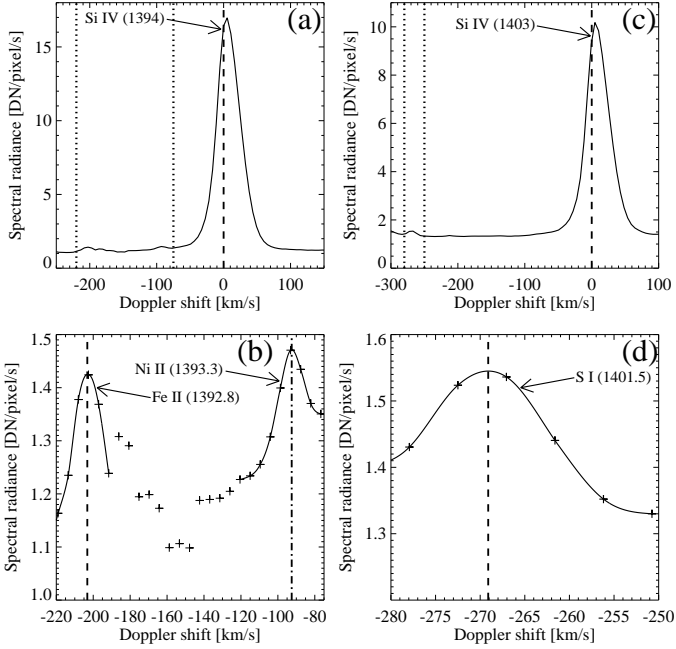}
   \caption{Line profiles used for wavelength calibration
 averaged over the whole FOV of the raster map. Panels (a) and (b) show
   the profile during the second raster. The main target line Si\,IV at 1394\,\AA\
   is visible in (a). The vertical dotted lines indicate the wavelength region shown
   in panel (b), where we zoom in on the  lines of Ni\,II and Fe\,II that were used for
   wavelength calibration.  The pluses show the actual observed spectrum,
   the solid line a spline fit around the center of the calibration
   lines. The vertical dashed lines indicate the peak of the spline fit.
   Panels (c) and (d) show the same for the third raster scan in the
   self-helicity phase. Here the
   S\,I line was used for wavelength calibration and the main target
   line was Si\,IV at 1403\,\AA.
   The count rate is per 0.166\arcsec$\times$0.166\arcsec pixel.
   }
   \label{F:lineprofiles}%
\end{figure}

For the Doppler maps a wavelength calibration is required, and we
used the average spectra of the full FOV of each raster scan for
this. Figure \ref{F:lineprofiles} shows the average spectra  for the
region around the 1394\,\AA\ (a,b) and 1403\,\AA\ (c,d)\ lines in
the second and third raster scans. The lower panels show the
(weaker) calibration lines alone. In the first and second scans
these are the Ni\,II and Fe\,II lines, in the third scan this is the
S\,I line (see Table \ref{table:2} for a list of these lines). In
part because of shortness of the scans, which lasted  for half an
hour or less, we verified that we did not have to correct the
Doppler shifts for the orbital motion of the spacecraft.

\begin{table}
\caption{Lines of interest. Listed are the rest wavelength, the Doppler
shift of the rest wavelength of the lines to the Si\,IV lines, and the
absolute Doppler shifts of the lines in the spectrum averaged over the
whole field of view after the wavelength calibration. Positive values
correspond to redshifts.}             % title of Table
\label{table:2}      % is used to refer this table in the text
\centering                          % used for centering table
\begin{tabular}{c c c c}        % centered columns (4 columns)
\hline\hline                 % inserts double horizontal lines
line & rest wavelength [\AA] & $\Delta$v [km/s] & Doppler shift [km/s] \\
    % table heading
\hline                        % inserts single horizontal line
Fe\,II & 1392.816 & -203.2 & 0.0 \\
Ni\,II & 1393.33  & -92.5  & 0.0 \\
Si\,IV (1394) & 1393.76 & 0 & +3.9 \\ \hline
S\,I  & 1401.163  & -269.1 & 0.0 \\
Si\,IV (1403) & 1402.773 & 0 & +5.9 \\
\hline                                   %inserts single line
\end{tabular}
\end{table}

\section{Mutual helicity of two loops}

During the first and the second raster scans, two cool loops
crossing each other are visible in the IRIS 1400\,\AA\ SJI
(Figs.\,\ref{F:images}a,b). The crossing point is marked by green
arrows in Figs.\,\ref{F:images}a and \ref{F:images}b. These loops
show emission from the Si\,IV lines and not the continuum that also
contributes to the SJI. This is clearly evident  from comparing
these SJI with the spectroheliograms in Si\,IV that were derived
from the spectral profiles of Si\,IV. These loops show plasma at
temperatures of about 10$^5$\,K. They connect two plage-type faculae
of the active region with opposite polarities (cf. the magnetogram
in Fig.\,\ref{F:images}e) and have lengths of about 40 Mm.

According to earlier work (Berger \cite{berger99}; R\'egnier et al.
\cite{regnier05}), these two loops should have mutual helicity
because they cross each other. Therefore, we call the time span
covered by the first two scans with the crossing loops \emph{the
mutual-helicity phase} below.

To better compare the AIA 304\,\AA\ images we also overplot the
envelope of the two loops as derived from the IRIS 1400\,\AA\ image
(Fig.\,\ref{F:images}a) on top of the AIA image
(Fig.\,\ref{F:images}d, yellow dotted lines). In principle, the
He\,II line dominating the AIA 304\,\AA\ channel and the Si\,IV
lines form at similar temperatures just below 10$^5$\,K. However,
the AIA 304\,\AA\ channel is also very sensitive to cool material
that is absorbed in the Ly-continua of H\,I and He\,I. This is contrast to
the 1400\,\AA\ emission in IRIS images because this is longward of
the Ly edge at 911\,\AA. This is the reason why the AIA 304 \,\AA\
image looks  quite different from the IRIS 1400\,\AA\ channel: while
the latter shows the cool loop, in the AIA 304\,\AA\ channel a dark
mini-filamentary structure is visible that highlights the cool material
caught on fieldlines next to the Si\,IV loops. At least in the
early phase of the first raster scan, there is no signature
in the 171\,\AA\ channel of AIA, which underlines that the loops seen in
IRIS are indeed cool loops. Later the 171\,\AA\ channel shows loops
alongside the cool IRIS loops, which indicates heating
of the loop system (see movie attached to Fig.\,\ref{F:images}).

\begin{figure}
   \centering
%   \sidecaption
   \includegraphics[width=8.8cm]{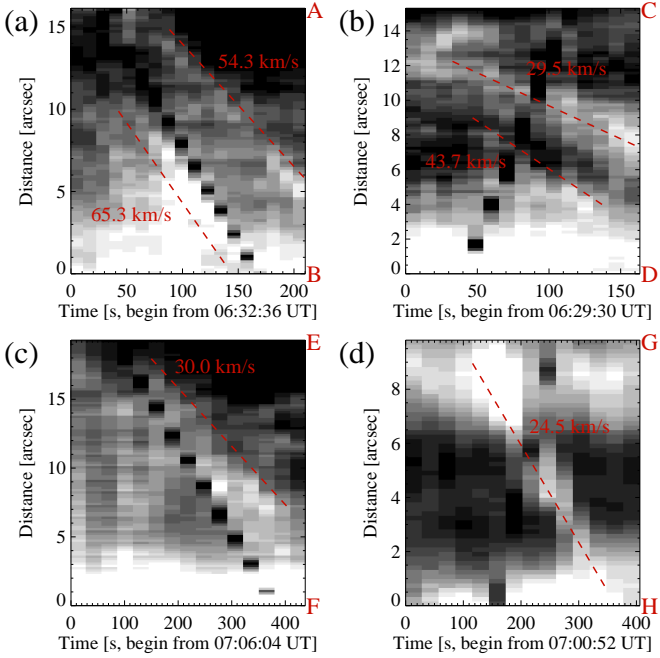}
   \caption{Proper motions along the Si\,IV loops.
   Time-space plots of a series of IRIS slit-jaw images at 1400\,\AA\ along
   the dashed lines AB (a), CD (b), EF (c), and GH (d), as marked in
   Figs.\,\ref{F:images}b and \ref{F:images}c. The red dashed lines show
   motions along the loops.
   The respective velocities are denoted by the numbers in the plots.}
   \label{F:timeslices}%
\end{figure}

In this mutual-helicity phase (during the first two raster scans),
apparent motions along the loops can be seen in the movies of the
IRIS 1400\,\AA\ SJI. Like leaves in the wind, bright blobs stream
along the loops. In Figs.\,\ref{F:timeslices}a and b we show
time-space plots of IRIS 1400\,\AA\ SJI during the second raster
scan (along lines AB and CD in Fig.\,\ref{F:images}b). We see
several blobs moving toward the footpoints of the loops (at B and D)
with apparent motions ranging from 30 km s$^{-1}$ to 65 km s$^{-1}$.

\subsection{Motion and interaction of two loops}

To investigate the motion of the two loops we studied the parameters
of the (single-) Gaussians fitted to the line profile of Si\,IV.
They are shown in Fig. 4 for the second raster scan, that is, in the
mutual-helicity phase. The intensity map shows the same cool loops
as the slit-jaw image (Fig.\,\ref{F:images}b). The loop crossing
occurs near the line marked f.

The Doppler map in Fig.\,\ref{F:spectrographs2}b reveals the general
motion of the two loops. Left of the crossing point, loop\,1 shows a
redshift and loop\,2 a blueshift. On the right side, loop\,2 now
shows a redshift and loop\,1 a slight blueshift. For a better view,
we display the spectra of Si\,IV at 1394\,\AA\ in
Fig.\,\ref{F:spectra2} along the spatial locations indicated by the
vertical dashed lines a-i in Fig.\,\ref{F:spectrographs2}. The red
markers in Fig.\,\ref{F:spectra2} denote the positions of loop\,1,
the green markers the positions of loop\,2. At all these cuts,
except for the cut at the crossing point, f, the two loops can be
clearly distinguished in the spectra. This more detailed look at the
spectra (which are not strictly single-Gaussian as assumed for the
fits) confirms the above description of the (line-of-sight) motion
of the loops.

Most importantly, there are several faint randomly distributed
short-lived signatures of internal helical motions for both loops
(spectral tilts in Figs.\,\ref{F:spectra2}c, \ref{F:spectra2}d, and
\ref{F:spectra2}h for loop\,1, and Figs.\,\ref{F:spectra2}a,
\ref{F:spectra2}b, and \ref{F:spectra2}h for loop\,2). The only
long-lived signature for a spectral tilt is found at the
crossing of the loop (cut f, Fig.\,\ref{F:spectra2}f). This is also
the only place where the Si\,IV line becomes very broad, which is very
clear from the line width plot in Fig.\,\ref{F:spectrographs2}c. At
other locations along the loop, the line width is typically only
some 15 km\,s$^{-1}$ to 20 km\,s$^{-1}$, while near the crossing
point it is twice as broad (cf. the arrow in
Fig.\,\ref{F:spectrographs2}c). Therefore the spectral tilt at
location f might be more an indication of the (small-scale) activity
due to the interaction of the two loops than of a helical motion.

We also checked the profiles  of the Mg\,II k line at 2796\,\AA\
that originates from chromospheric plasma. In the appendix
Figs.\,\ref{appintensity1} and \ref{appspectra1} show the two loops
during the second scan in the mutual-helicity phase. Even though the
signal from Mg\,II is much fainter, we obtain similar results as for
the Si\,IV line (see more details in Appendix A).

\subsection{Scenario for loop interaction}

The Doppler motions of the loops in the mutual-helicity phase
indicate that the two loops wind around each other, interlocked near
the crossing point. At the crossing point the magnetic fields hosting
the two loops interact, which could lead to reconnection at that
place. On the one hand, this would explain the increased line width
seen at the crossing point, which would be a signature of the heating
process that accompanies reconnection. On the other hand, the
reconnection would accelerate material along the field lines, hence
the loops, away from the crossing points in the direction of the
footpoints, which would be consistent with the apparent motion seen
in the slit-jaw images discussed above. Another consequence of that
heating process are loops in the 171\,\AA\ channel of AIA
in the later mutual-helicity phase (see movie attached
to Fig.\,\ref{F:images}). These indicate hotter plasma that is found
in loops next to the cool loops of the IRIS slit-jaw images.

In brief, these observations are consistent with a scenario in
which the two loops
have mutual helicity and wind around each other, followed by
reconnection at the crossing point.

\section{Conversion to self-helicity}
\begin{figure}
   \centering
   \includegraphics[width=8.8cm]{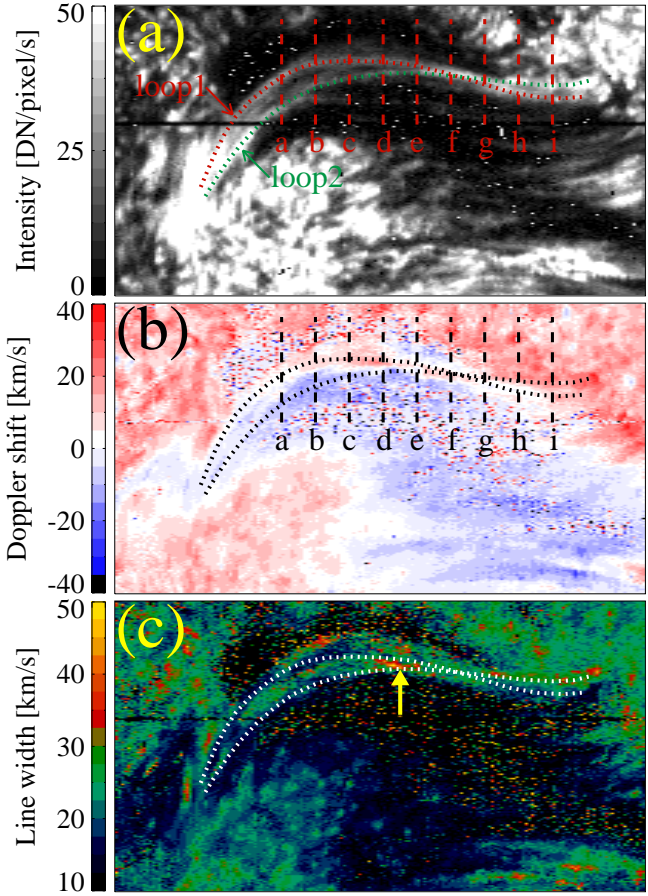}
   \caption{Maps of intensity (a), Doppler shift (b), and line width (c)
   of Si\,IV (1394\,\AA) for the second raster scan in
   the mutual-helicity phase.
   The dotted lines highlight the location of the two cool loops. The dashed
   lines a-i in (a) and (b) indicate the positions of the spectra
   shown in Fig.\,\ref{F:spectra2}. The yellow arrow in (c) marks
   a region with enhanced line width. The location of the maps is indicated
   in Fig.\,\ref{F:images}b. The FOV is 60.5$\arcsec$$\times$33.3$\arcsec$.}
   \label{F:spectrographs2}%
\end{figure}

\begin{figure}
   \centering
   \includegraphics[width=8.8cm]{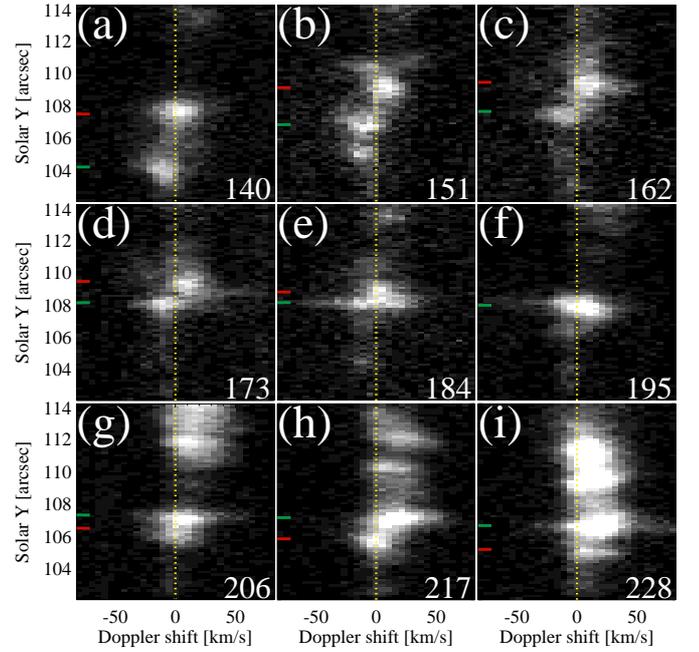}
   \caption{Line profiles of Si\,IV (1394\,\AA) during the
   mutual-helicity phase. The spatial location of these spectra is
   indicated in Fig.\,\ref{F:spectrographs2} by the vertical dashed lines a-i. The red and
   green markers show the location
   of the two loops (red and green dotted lines in Fig.\,\ref{F:spectrographs2}a). The numbers
   in the plot indicate the raster index, the vertical dotted lines the
   zero Doppler shift.}
   \label{F:spectra2}%
\end{figure}

\subsection{Loops after interaction}

After the interaction of the two loops, the situation changes quite
dramatically, as can be seen in Fig.\,\ref{F:images}c, which
displays an IRIS 1400\,\AA\ SJI during the third raster scan. As we
show below, this phase shows signatures of self-helicity, which is
why we call this \emph{the self-helicity phase}.

After the interaction, one of the two loops remains visible in full
length (labeled loop\,1 in the figures). However, the middle part of
the other loop disappears (loop\,2).  By combining the simultaneous
AIA/SDO 171\,\AA\ images, we found that these two Si\,IV loops
basically run  parallel after the interaction. In particular, the
Si\,IV loops are now also visible in the 171\,\AA\ channel of AIA
(Fig.\,\ref{F:images}f), which is different from the earlier
mutual-helicity phase. The 171\,\AA\ channel mainly shows plasma
just below 10$^6$\,K. That the loops in the AIA 171\,\AA\ images are
roughly co-spatial with the cool Si\,IV loops indicates that (part
of) the plasma is heated in response to the reconnection
process.\footnote{We cannot fully rule out that the emission in the
AIA 171\,\AA\ band is due to cool plasma because this channel shows
some contamination from O\,V lines. However, based on the temporal
evolution we do not consider this likely, because then the early
part of the mutual-helicity phase of the 171\,\AA\ channel should
have shown the loops when they were already clearly visible in the
transition region emission in Si\,IV.} The temporal evolution of the
AIA 171\,\AA\ emission is shown in the movie attached to
Fig.\,\ref{F:images}.

Similar to the mutual-helicity phase, bright blobs also move along
the loops during the self-helicity phase. Figures
\ref{F:timeslices}c and \ref{F:timeslices}d show two time-space
diagrams of a series of 1400\,\AA\ slit-jaw images along the dashed
red lines EF and GH denoted in Fig.\,\ref{F:images}c. The proper
motions away from the (former) reconnection site remain. The
velocities are now lower by almost a factor of two and range between
25 km s$^{-1}$ and 30 km s$^{-1}$.

\begin{figure}
   \centering
   \includegraphics[width=8.8cm]{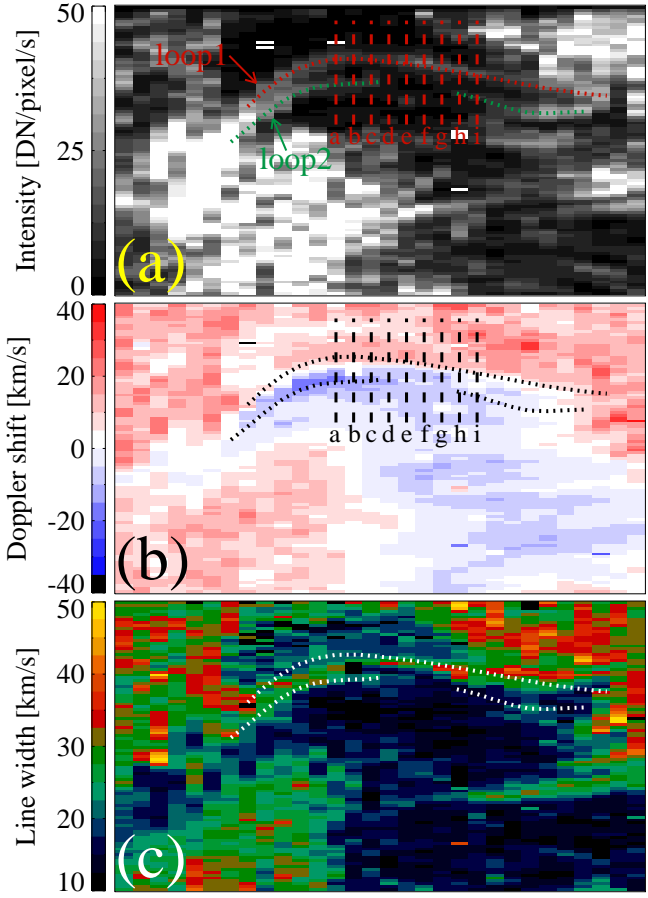}
   \caption{Similar to Fig.\,\ref{F:spectrographs2}, but for the third raster scan in the self-helicity phase. The dotted lines highlight the Si\,IV loops, the vertical
   dashed lines indicate the position of the spectra, now shown in Fig.\,\ref{F:spectra3}.
   Same FOV as in Fig.\,\ref{F:spectrographs2}.}
   \label{F:spectrographs3}%
\end{figure}

The maps of intensity, Doppler shift, and line width of the Si\,IV
line at 1403\,\AA\ for the third raster scan, that is, the self-helicity
phase, are displayed in Fig.\,\ref{F:spectrographs3}. As  seen from
the slit-jaw images, in this spectroheliogram the (former) two
loops no longer wind around each other here either, but run in parallel, with
loop\,2 scarcely visible in the middle part. The intensity
of Mg\,II also shows similar loop structures (see Appendix A for
details). The line width of the Si\,IV profile along the two loops
is narrow, similar to that of the mutual-helicity phase (outside the
reconnection/crossing point).

\subsection{Self-helicity in a loop}

The most significant change from the earlier mutual-helicity phase
is found in the spectral profiles of Si\,IV. To illustrate this, we
show in Fig.\,\ref{F:spectra3} the spectral images at several slit
positions, labeled a to i in Fig.\,\ref{F:spectrographs3}. In the
middle part of loop\,1 (red markers) the Si\,IV line shows a clear
spectral tilt from positions e through h
(Figs.\,\ref{F:spectra3}e-f). Toward the left and right footpoints
the spectra show no such spectral tilts.

The spectral tilt in the middle part of the loop is very clear, and
the centroid of the profile changes from red to blue across a few
pixels within the clearly defined cool loop. From
Figs.\,\ref{F:spectra3}e to \ref{F:spectra3}h, one can estimate the
spectral tilt from about 10 km s$^{-1}$ blueshift on the southern
side to about 30 km s$^{-1}$ on the northern side. This is a
signature of internal helical motions and indicates that loop\,1 now
has self-helicity (e.g., the twist). The profiles of the Mg\,II line
show a similar spectral tilt, but here with weaker Doppler shifts
than in Si IV (see Appendix A and Fig.\,\ref{appspectra2} for
details).

In summary, this is evidence that the mutual helicity of the
two
loops turned into
the internal helicity of a single loop.

\subsection{Pitch angle of the flow}

\begin{figure}
   \centering
   \includegraphics[width=8.8cm]{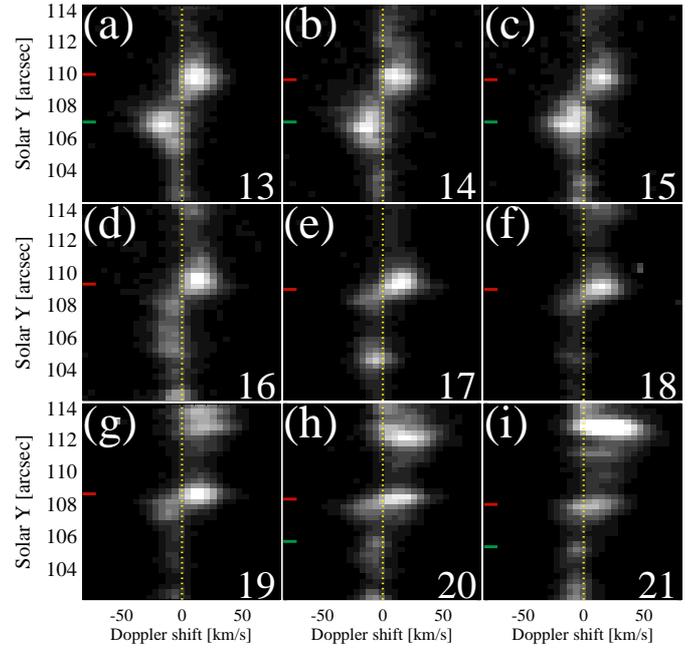}
   \caption{Similar to Fig.\,\ref{F:spectra2}, but for the self-helicity phase, now showing
   Si\,IV (1403\,\AA). The red and green markers show the position
   of the loops indicated
   by red and green dotted lines in Fig.\,\ref{F:spectrographs3}a.}
   \label{F:spectra3}%
\end{figure}

From the images and spectra observed by IRIS, we can estimate the
real plasma motions along loop\,1 during the self-helicity phase.
The apparent velocity of the proper motion along loop\,1 in the
self-helicity phase is about 30 km s$^{-1}$
(Fig.\,\ref{F:timeslices}c and Sect. 4.1). The spectral tilt across
loop\,1 is of about 40 km s$^{-1}$ (Figs.\,\ref{F:spectra3}e to
\ref{F:spectra3}h and Sect. 4.2). From this one can estimate that
the pitch angle of the internal helical motion is of about
20$^{\circ}$ to 45$^{\circ}$. Consequently, the speed of the helical
motion is about 40 km s$^{-1}$, which is close to the sound speed of
about 50 km s$^{-1}$ near 100 000 K in the line formation region of
Si\,IV.

\section{Discussion and conclusions}

In an observation with IRIS we found two cool loops with mutual
helicity that interact. In this process the mutual helicity is
transformed into self-helicity. Images acquired with AIA/SDO support
that cool structures are present in the mutual-helicity phase, which
are then heated by the interaction of the loops. The spectroscopic
observations with IRIS clarify that the cool loops wind around each
other and produce an increased line broadening due to the
interaction and heating at the location of the loop crossing. The
spectral tilt of the spectra within the single remaining full cool
loop is a clear indication for the conversion of mutual into
self-helicity. The pitch angle of the final internal helical motion
in that loop is about 20$^{\circ}$ to 45$^{\circ}$, the average
velocity is about  30 km s$^{-1}$ to 45 km s$^{-1}$. The
filament-type structures in the 304\,\AA\ channel of AIA are
consistent with the mutual and self-helicity in filaments (Chae
\cite{chae00}). The bright dots in the AIA 171\,\AA\ channel might
be indicative of hot plasma that is driven away from the interaction
site.

Our investigation of the loops in Mg\,II\,k showed results similar
to those for Si\,IV, albeit less clearly. This implies that these
loops are multithermal: First, the Mg\,II intensity shows loops that
are roughly co-spatial with those in Si\,IV. Second, and more
important, the spectral profiles of Mg\,II and Si\,IV show similar
characteristics at the location of the loops. Therefore it is highly
unlikely that what we see in Mg\,\,II is just some low-lying plasma
along the line of sight. Instead, the source region of Mg\,II has to
be the same volume as the source region of Si\,IV. Thus the loops we
studied contain chromospheric \emph{and} transition-region plasma.

Tilts of spectral lines have been observed before and have been
interpreted as rotational or twisting motions (e.g., Pasachoff et
al. \cite{pasachoff68}; Rompolt \cite{rompolt75}; Curdt et al.
\cite{curdt12}; De Pontieu et al. \cite{depontieu12}). Recent IRIS
observations showed that these twisting motions are a ubiquitous
phenomenon on the Sun (De Pontieu et al. \cite{depontieu14b}). The
random faint short-lived spectral tilts during the mutual-helicity
phase are similar to the ubiquitous propagating twists along
chromospheric features reported by De Pontieu et al.
(\cite{depontieu14b}), which are signatures of self-helicity-driven
spicule-like features. It has been proposed before that the
conversion from mutual  into self-helicity might occur when two
crossing loops interact through reconnection (Berger
\cite{berger99}, Georgoulis \cite{georgoulis11}, Tziotziou et al.
\cite{tziotziou13}). However, direct observational evidence for this
conversion process could not be found. The new IRIS observations
presented here allow following this conversion process and thus
provide a new insight into how loops with internal twists might
form. In the self-helicity phase the spectral tilt is distributed
along the loop and is not only concentrated at the crossing point of
the two loops. This indicates that the self-helicity propagates away
from the magnetic reconnection site. Unfortunately, the current data
do not allow following the temporal evolution of the self-helicity
signatures, which one might naively expect to propagate with the
Alfv\'en speed.

\begin{figure}
   \centering
   \includegraphics[width=8.8cm]{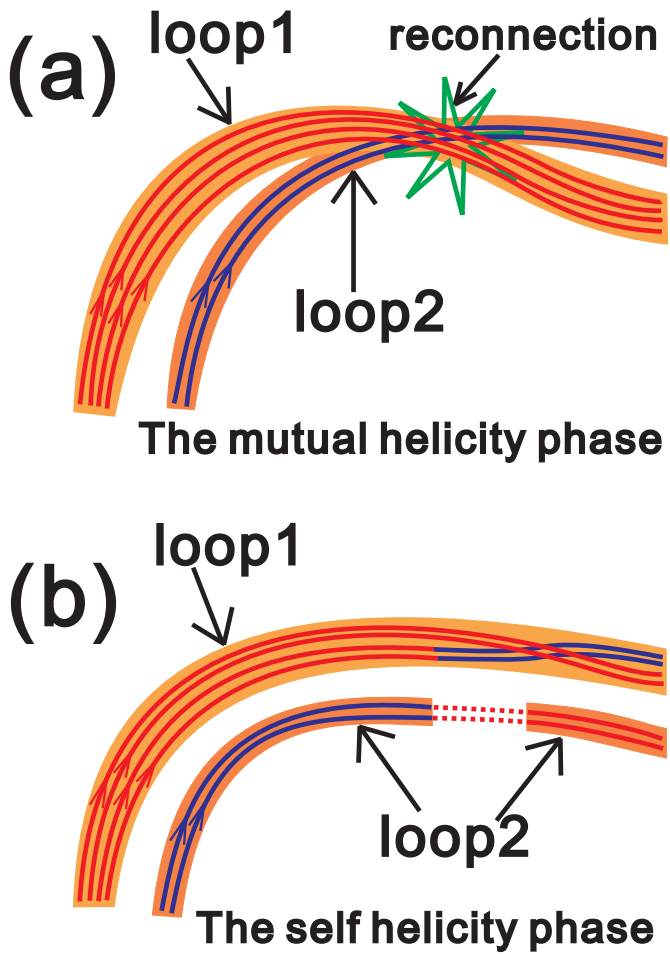}
   \caption{Schematic diagrams illustrating the configurations of two
   loops, the orange-colored broad lines of loop\,1 and loop\,2, during
   the mutual helicity (a) and the self-helicity phases (b). The red lines in (a) show the magnetic field lines
   of loop\,1, the blue lines show the field lines of
loop\,2. The green star denotes magnetic
   reconnection. The red-blue lines in (b) indicate the reconnected lines,the dotted red lines show loops that are not detected in Si\,IV images.
   The red and blue arrows show the directions of the magnetic field lines.}
   \label{F:diagrams}%
\end{figure}

The scenario for the conversion of mutual into self-helicity is
sketched in Fig.\,\ref{F:diagrams}. Here we interpret the conversion
as interchange reconnection between the two crossing loops. The blue
and red lines show the magnetic field lines, the more diffuse
orange-colored broad lines represent the actual cool loops as seen
in the 1400\,\AA\ channel of IRIS. In the mutual-helicity phase the
red fieldlines fully belong to loop\,1, the blue fieldlines fully to
loop\,2 (Fig.\,\ref{F:diagrams}a): The loops wind around each other,
which is the same as to say as that they have mutual helicity. At
the crossing point reconnection sets in (green star), which marks
the beginning of the end of the mutual-helicity phase. After the
reconnection there are  again two loops, but now more or less
running in parallel with some of the fieldlines exchanged. The upper
of the two loops in this later stage is now internally twisted
within itself, which is a consequence of the interchange
reconnection: now loop\,1 has self-helicity. Flows are initiated
during the reconnection process, and as they move along the loop
with self-helicity, they show a twisting motion, which is detectable
in the observations as a spectral tilt. In the end, this schematic
diagram is similar to the process in which helical flux ropes are
produced by reconnection (e.g., Moore et al. \cite{moore01}; Cheng
et al. \cite{cheng14}). Using Hi-C observations, Cirtain et al.
(\cite{cirt13}) showed evidence of braiding, subsequent
reconnection, and straightening of the loops in the corona. Their
scenario is similar to what we displayed in Fig.\,\ref{F:diagrams},
but for multiset loops. However, our additional spectral data enable
us to identify line-of-sight motions of, and even the motions
inside, the unresolved loop strands. The IRIS observations presented
here show a direct observational signature of this scenario of a
conversion of mutual into self-helicity.

\begin{acknowledgements}

IRIS is a NASA Small Explorer mission developed and operated by
LMSAL with mission operations executed at NASA Ames Research center
and major contributions to downlink communications funded by the
Norwegian Space Center (NSC, Norway) through an ESA PRODEX contract.
This work is supported by NASA contract NNG09FA40C (IRIS), the
Lockheed Martin Independent Research Program, the European Research
Council grant agreement No. 291058 and NASA grant NNX11AO98G.
The AIA and HMI data used are provided courtesy of NASA/SDO and the
AIA and HMI science teams.
This work is supported by the National Basic Research Program of
China under grant 2011CB811403, the National Natural Science
Foundations of China (11303050, 11025315, 11221063) and the CAS
Project KJCX2-EW-T07.

\end{acknowledgements}

\Online
\clearpage

\begin{appendix} %First online appendix

\section{Chromospheric observations of mutual- and self-helicity in the two loops}

To relate the transition region plasma to
emission from chromospheric origin, we analyzed the Mg\,II\,k
(2796\,\AA) line of these two loops observed by IRIS in the same way as for the Si\,IV data shown in the main part of
this study.

\begin{figure}
\centering
\includegraphics[width=8.8cm]{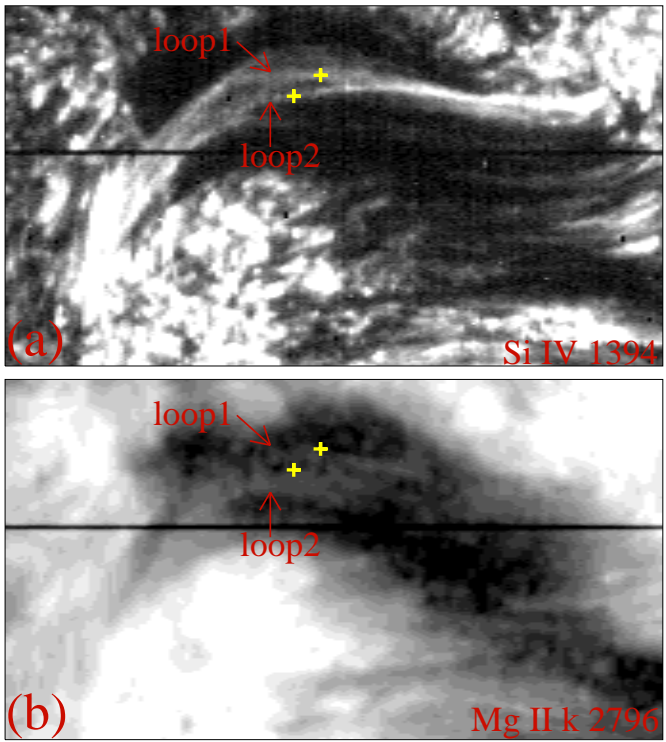}
\caption{Intensity maps of IRIS Si\,IV (1394\,\AA) (a) and
Mg\,II\,k (2796\,\AA) (b) during the second scan in the mutual-helicity phase. The yellow pluses mark the positions of the Si\,IV
loops. The FOV is the same as
in Fig.\,\ref{F:spectrographs2}a.} \label{appintensity1}
\end{figure}

Figure\,\ref{appintensity1} displays the loops during the second
scan in the \emph{mutual-helicity phase}.
Figure\,\ref{appintensity1}b shows the intensity of the Mg\,II line.
Two fainter loops are detected (roughly) co-spatially with the
Si\,IV loops, as displayed in Fig.\,\ref{appintensity1}a (see also
Fig.\,\ref{F:spectrographs2}a). This applies especially to the
region surrounding the middle part of these two loops, which
is marked by two
red arrows.

The yellows crosses in the two panels of Fig.\,\ref{appintensity1}
are at exactly the same positions. The alignment between
the Si\,IV and Mg\,II maps is easily achieved using the    fiducial
marks on the slit that are visible in the spectro-heliograms as
horizontal black lines. The yellow crosses mark two positions along
the Si\,IV loops, and a comparison of panel a and b of
Fig.\,\ref{appintensity1} shows that the loops of the Mg\,II k line
are located $\sim$1\arcsec\,to the south of the Si\,IV loops. Thus
the chromospheric and transition region plasma is not at exactly the
same position, but they most likely populate different strands of
the same loop system.

\begin{figure}
\centering
\includegraphics[width=8.8cm]{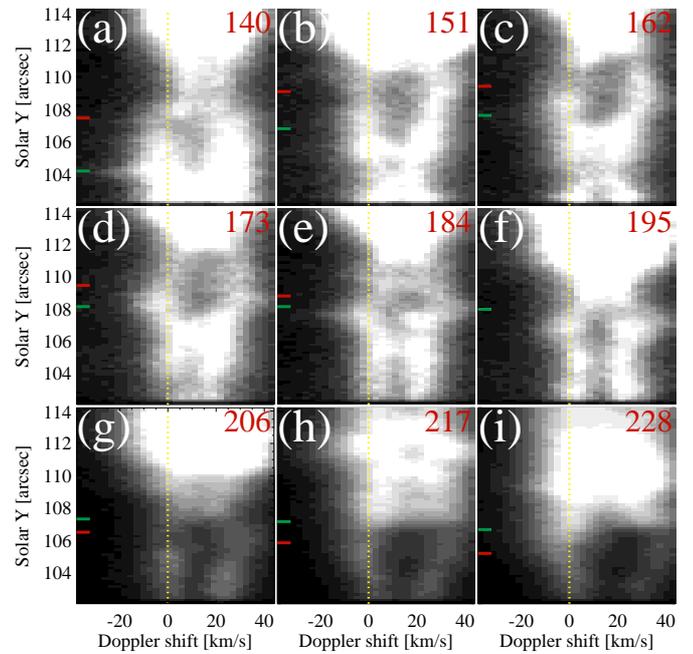}
\caption{Similar as Fig.\,\ref{F:spectra2}, but for the Mg\,II\,k
(2796\,\AA) line.} \label{appspectra1}
\end{figure}

Figure\,\ref{appspectra1} displays the profiles of Mg\,II k at
2796\,\AA\ in the mutual-helicity phase. The positions of the
spectra are the same as those of the Si\,IV profiles shown in
Fig.\,\ref{F:spectra2}. The red and green marks here indicate the
positions of the Si\,IV loops as displayed in
Fig.\,\ref{F:spectra2}. By comparing Fig.\,\ref{appspectra1} with
Fig.\,\ref{F:spectra2}, we can find some similar indications of the
Doppler motions of the loops, even though the signals are weaker in
the Mg\,II k line.

\begin{figure}
\centering
\includegraphics[width=8.8cm]{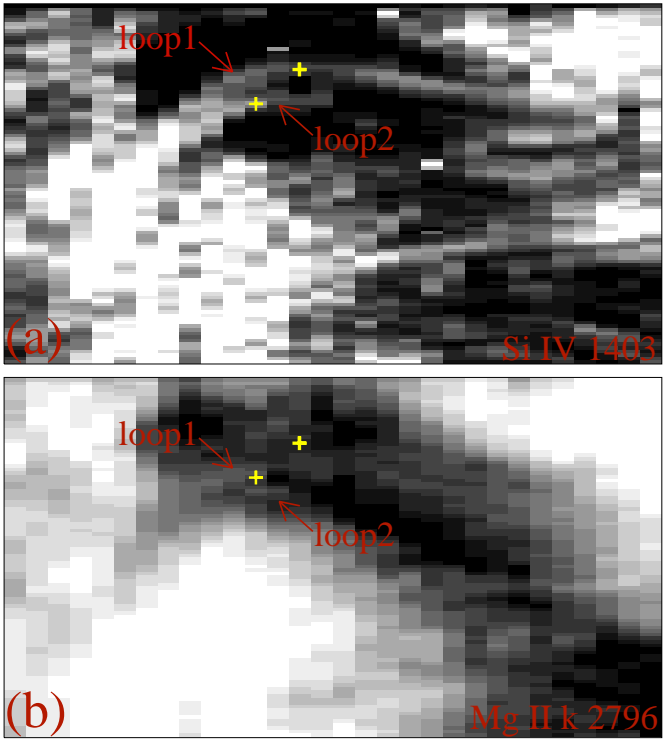}
\caption{Same as Fig.\,\ref{appintensity1}, but for the self-helicity
phase, now showing the Si\,IV (1403\,\AA) (a) and Mg\,II k
(2796\,\AA) line (b).} \label{appintensity2}
\end{figure}

Figure\,\ref{appintensity2} shows the loops during the third scan in
the \emph{self-helicity phase}. Here, again, the Mg\,II line shows
roughly the same loop patterns. Just as in the earlier
mutual-helicity phase, here the loops seen in Mg\,II seem to be
offset by $\sim$1\arcsec\ to the south of the Si\,IV loops (because
the FOV is different in this raster-scan map, the fiducial mark
is not visible in Fig.\,\ref{appintensity2}).

\begin{figure}
\centering
\includegraphics[width=8.8cm]{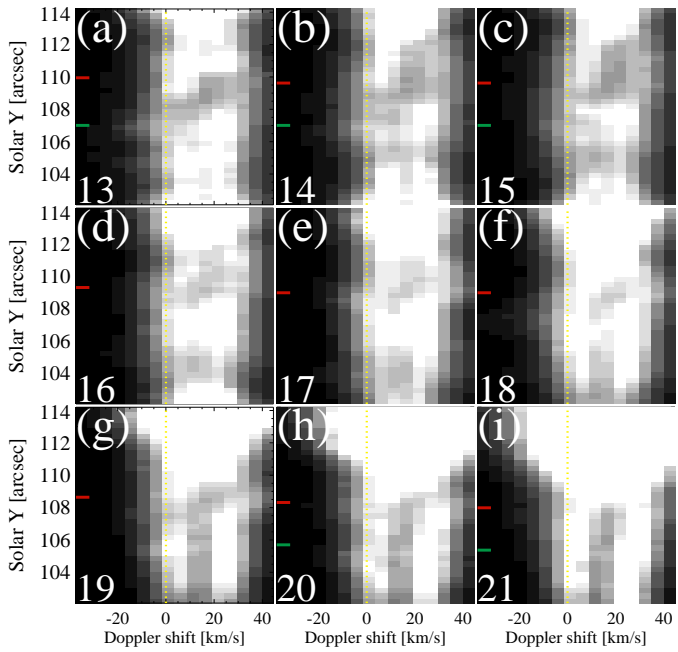}
\caption{Same as Fig.\,\ref{appspectra1}, but for the self-helicity
phase.} \label{appspectra2}
\end{figure}

In Fig.\,\ref{appspectra2} we show  the spectra of Mg\,II now in the
self-helicity phase, again at the same positions as those of Si\,IV
in the main text (cf. Fig.\,\ref{F:spectra3}). For the north loop,
loop\,1, some spectral tilt is detected (see
Figs.\,\ref{appspectra2}e-\ref{appspectra2}h). These spectral tilts
are similar to those displayed in Fig.\,\ref{F:spectra3} for Si\,IV,
but with weaker Doppler shifts.

From these considerations we conclude that the spectro-heliograms in
Mg\,II show loop structures similar to those in Si\,IV, albeit
offset by about 1\arcsec\ perpendicular to the loop spine. Most
importantly, even the spectral profiles of Mg\,II and Si\,IV share
similar properties. This implies that the source region of the
chromospheric Mg\,II\,k line and the transition region line of
Si\,IV share the same volume defined by the loop system. Thus the
loops are indeed multithermal structures, probably with the
chromospheric and transition region plasma found on different
strands within the loop system. However, because the signals in the
Mg\,II k line are much weaker than in the Si\,IV line, the results
we discussed here need to be considered carefully.

\end{appendix}

\end{document}